\documentclass[twocolumn,final]{elsart5p}
\usepackage{graphicx}
\usepackage{color}
\journal{Nuclear Instruments and Methods B}
\begin{document}
\begin{frontmatter}
\title{HICS: Highly charged ion collisions with surfaces}
\author{T. Peters},
\author{C. Haake},
\author{J. Hopster},
\author{V. Sokolovsky},
\author{A. Wucher},
\author{M. Schleberger\corauthref{cor}}
\corauth[cor]{Corresponding author. {\it Email adress:} {\bf marika.schleberger@uni-due.de}}

\address{Experimentelle Physik, Universit\"at Duisburg-Essen, 47048 Duisburg, Germany}

\begin{abstract}
The layout of a new instrument designed to study the interaction of highly charged ions with surfaces, which consists of an ion source,  a beamline including charge separation and a target chamber, is presented here.
By varying the charge state and impact velocity of the projectiles separately, the dissipation of potential and kinetic energy at or below the surface can be studied independently.
The target chamber offers the use of tunable metal-insulator-metal devices as detectors for internal electronic excitation, a time-of-flight system to study the impact induced particle emission and the possibility to transfer samples {\em in situ} to a UHV scanning probe microscope. Samples and detectors can be prepared {\em in situ} as well. As a first example data on graphene layers on SrTiO$_3$ which have been irradiated with Xe$^{36+}$ are presented.
\end{abstract}

\end{frontmatter}

\section{Introduction}

The interaction between energetic ions and solid surfaces may be studied in many different ways. To get the complete picture one has to look upon several different energy dissipation mechanisms. The kinetic energy of the projectile is transferred by either elastic collisions (nuclear stopping) or by electronic stopping, while its potential energy can be transferred by electron capture, Auger and radiative transitions. The energy dissipation manifests itself e.g.~in
the emission of 
sputtered particles (ions, neutrals, clusters) \cite{sputterbuch}, in the emission of electrons \cite{Zwart1989,Elektronenemission} or in the creation of permanent modifications
at the surface 
\cite{SHIreview,NatNano,Fritzreview}. If slow highly charged ions are used as projectiles, a wide range of these processes can be studied \cite{HCIreview} without the need of a large scale accelerator facility.

\begin{figure}
\includegraphics[width=\columnwidth]{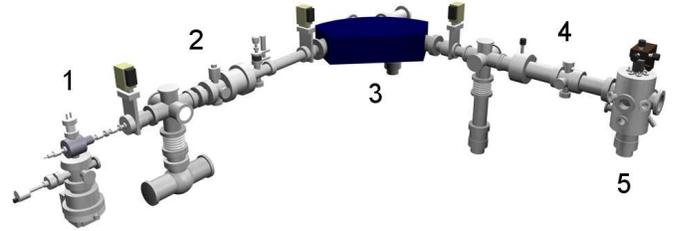}
\caption{Schematic view of the HICS setup. (1) EBIT, (2) beamline, (3) sector magnet, (4) ion escalator, (5) target chamber. The various flanges of the target chamber allow the mounting of a load lock, a preparation chamber, a ToF mass spectrometer and the laser system.}
\label{HICS}
\end{figure}

In order to investigate the interaction of slow highly charged ions with surfaces a new ultra high vacuum (UHV) system with the name HICS (highly charged ions collisions with surfaces) has been constructed. HICS consists of three major parts, the ion source (1), the beamline (2) including charge separation (3) and ion deceleration (4), and the target chamber (5) including a load lock system (6) for preparation and target transfer (7) as shown in fig.~\ref{HICS}.

\section{HICS}
To produce highly charged ions a Dresden EBIT (Electron Beam Ion Trap, \cite{Werner2001}, from DREEBIT GmbH) is mounted on the HICS setup. The operation principle is based on ionisation by successive electron collisions while trapping the ions in an electric and magnetic field \cite{Marrs1988}. The created charge state distribution depends on the time the ions are trapped within the drift tube and can be checked by detecting the x-ray emission spectrum (AmpTek XR-100CR detector). When opening the trap by lowering the voltage of the front electrode, a short ion pulse is released into the beamline. A ``leaky'' operation mode, where the trap is not completely closed and produces a continuous beam, is also possible,
but the creation of highly charged ions is less effective due to the shorter time the ions remain in the trap.
As projectiles noble gases like argon (up to Ar$^{18+}$) and xenon (up to Xe$^{46+}$) are used.
The gas pressure during operation is typically between $1 \cdot 10^{-10}$ and $6~\cdot 10^{-9}$~mbar, with a base pressure below $10^{-10}$~mbar.
The drift tubes of the ion trap have a potential between 5 and 15~kV with reference to the beamline and the trap depth is typically 100~V.
To obtain a well defined charge state from the distribution the ion pulse is led into a bending sector magnet (Danfysik). A measurement of the ion current in a Faraday cup behind the magnet (bending radius of 90$^{\circ }$) reveals the different ion charge states for a given magnetic field. Selecting a single charge state from this distribution can be done by choosing an appropriate magnetic field and limiting the beam width to a beam consisting of a single charge state by a variable slit.

As the drift tubes have an electric potential $U_{\mathrm D}$ of several kilovolts with reference to the beamline, the ions with charge state $q$ enter the beamline with a kinetic energy of $E_{\mathrm{kin}}=q \cdot U_{\mathrm D}$. The beamline is kept at ground potential for security and handling reasons and so is the irradiated sample. Therefore, a simple electric field between the beamline and the target cannot be used to decelerate the ions. Instead, a device which we call an ion escalator (IE) has been implemented. It consists of a 53~cm long metallic tube within the beamline, which is connected to a fast high voltage switch (Behlke HTS 151-03 GSM). The IE is positioned after the sector magnet to keep the ions at high velocity as long as possible.

\begin{figure} 
\includegraphics[width=8cm]{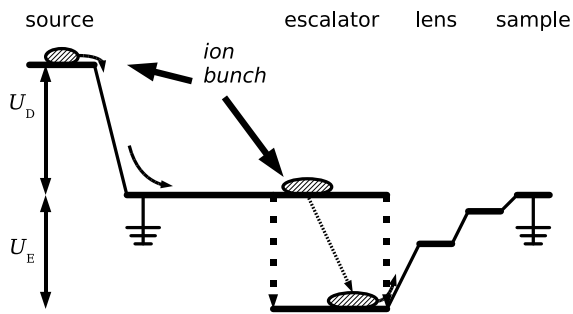}
\caption{Potential energy scheme of the ion escalator. As soon as the ion bunch has completely entered the tube of the IE, the tube is switched to a negative voltage with respect to the beamline. The ions are decelerated by the potential difference between tube and target after leaving the IE.}
\label{lift}
\end{figure}

A scheme of the operating principle can be seen in fig.~\ref{lift}. Before the ion bunch arrives at the IE it is at beamline potential.
Once the whole ion bunch has entered the tube the voltage is switched to a high negative potential, smaller than the value of the drift tube potential ($|U_\mathrm{E}| < |U_\mathrm{D}|$). At typical extraction voltages of a few kV, the length of the ion bunch is
shorter than the IE tube length. The switching is triggered by a programmable delay generator, which, in turn, is triggered by a signal generated by the opening of the ion source. Due to the fast rise time of approximately 20~ns, switching of the potential is completed before the ion bunch reaches the end of the tube. The variation of the potential has no effect on the velocity of the ions because the electric field in the interior of the tube remains zero. When the ions leave the IE they are decelerated by the remaining potential difference between the IE (now at negative potential) and the target (at ground potential). A 5-segment decelerating lens positioned between the end of the IE tube and the sample holder is used to focus the ion bunch onto the target. In summary, this setup allows the irradiation of targets with a constant kinetic energy as low as $E_{\mathrm{kin}} \approx$ 2000~eV independent of the charge state.

\section{Electronic excitations}
\begin{figure}
\includegraphics{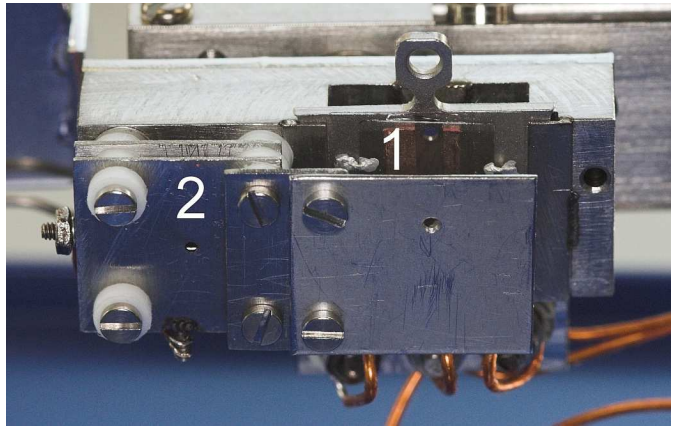}
\caption{Picture of the sample holder with a mounted MIM (1), consisting of two crossing metal films. A mask can be installed in front to prevent irradiation adjacent to the crossing area. A Faraday cup is installed next to the sample (2) for primary ion beam intensity measurements including secondary electron suppression.}
\label{FotoMIM}
\end{figure}

The target is mounted on a 5-axis manipulator in the irradiation chamber. To measure the primary ion current the manipulator is equipped with a Faraday cup.
Three metal pins are mounted on the sample holder to contact the surface of the sample, so that metal--insulator--metal (MIM, see fig.~\ref{FotoMIM}) junctions can be used to study electronic excitations below the surface \cite{Meyer2004,Peters2008,Kovacs2008}. The irradiated top metal layer is typically 25~nm thick, the insulator is about 3~nm thick. Excited electrons and holes
can tunnel from the top metal film into the bottom metal substrate,
where they can be detected as an internal emission current. By applying a bias voltage between the two metal electrodes the potential barrier can be varied so that the MIM acts as an energy dispersive element. MIM junctions consisting of silver on aluminum-oxide/aluminum have already been successfully used in previous studies \cite{Peters2008,Kovacs2008}. In the future, alternative junctions consisting of gold on tantalum-oxide/tantalum will be used, as well as different projectiles and higher charge states.

The setup is designed to be equipped with a time--of--flight (TOF) system \cite{Wucher1993a} for the detection of sputtered particles that is currently under construction. This mass spectrometer will encompass laser post-ionisation to allow the detection of secondary ions and neutrals. In this way, sputter yields as well as ionisation probabilities of sputtered particles can be measured as a function of the potential energy of the projectile, thus allowing to investigate the influence of electronic substrate excitations on the particle emission characteristics.

\section{Surface modifications}
Next to the irradiation chamber a sample preparation chamber is mounted where thin films can be evaporated onto samples by molecular beam epitaxy. In this way the thickness of the top layer of a MIM junction can be easily modified. The preparation chamber is equipped with a load lock system that allows for easy sample exchange as well as sample transfer to other UHV setups such as scanning probe microscopy (SPM) by means of a mobile vacuum system. Thus, it is possible to study ion induced surface modifications without breaking the vacuum. A direct mounting of the UHV-SPM to the HICS setup is not favourable due to mechanical vibrations from the beamline which would prevent atomic resolution. 

\begin{figure}
\includegraphics[width=\columnwidth]{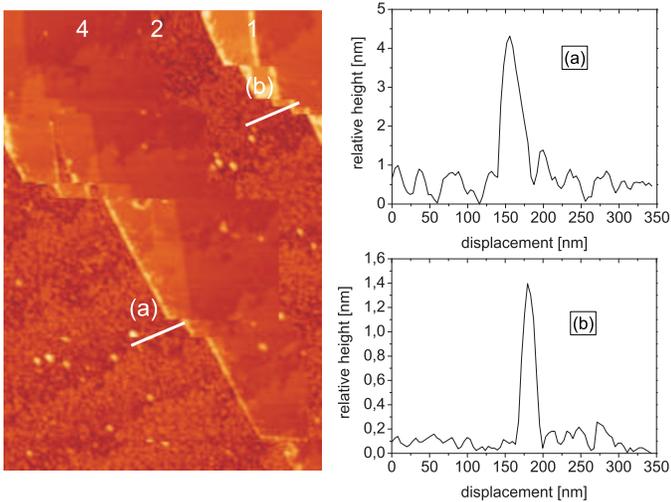}
\caption{AFM image (intermittent contact mode, NCHR-cantilever) of graphene on SrTiO$_3$ irradiated with Xe$^{36+}$. Frame size: 1.5~$\mu$m $\times$ 2~$\mu$m. Height scale see line scan. Single hillock like defect structures can be seen as bright spots. Line scan (a) shows an ion induced hillock on SrTiO$_3$ (lower left part of the image) . Line scan (b) shows a cross section of an ion induced hillock on a monolayer graphene (upper right part). The number of the graphene layers is given on top.}
\label{AFMBild}
\end{figure}

As a preliminary result, data that was obtained from a graphene sample that has been irradiated in the HICS setup, are presented here. The sample has been prepared {\em ex-situ} by mechanical exfoliation of highly oriented pyrolitic graphite \cite{Graphene} on the surface of a SrTiO$_3$(100) single crystal (Crystec, Berlin) \cite{SAkcoeltekin08}. This system represents a model system for a thin metal film (graphene) on an insulating substrate (SrTiO$_3$). It has been irradiated with 150 keV Xe$^{36+}$ ions applying a fluence of $1\cdot 10^{9}$ ions/cm$^2$. As finding graphene in a dedicated UHV AFM is very time consuming we chose to use an ambient AFM for demonstration purposes. In fig. \ref{AFMBild} one can clearly identify single nanosized hillocks protruding from the substrate surface. These hillocks appear to be identical to the ones found in experiments on CaF$_2$ \cite{ElSaid2008} and to the ones found in swift heavy ion irradiation experiments \cite{Khalfaoui2006,Akcoeltekin08}. This could be due to the fact that both types of irradiation induce predominantly strong electronic excitations which have been shown to lead to local melting processes. In the present work hillocks are also found on the surface areas covered with graphene. They appear significantly smaller and from a preliminary analysis they also seem to be less in number, depending on the number of graphene layers. For coverages of several graphene layers no defects could be found at all. The latter could be due to an extremely efficient energy dissipation process in the graphene layer compared to the insulating substrate. From a detailed analysis of this kind of experiments we hope to learn more about the physical processes related to highly charged ion impacts on such surfaces.

\section{Summary and Outlook}
The new HICS system presented here will allow detailed investigations of the ion--surface--interaction process. Potential and kinetic energy may be varied separately in a wide range including very high charge states at very low energies. Detection methods include a TOF mass spectroscopy system for the sputtered particles, metal--insulator--metal devices to study internal excitations as well as the possibility to study ion induced modifications {\em in situ} by means of scanning probe microscopy.
The preparation chamber allows
{\em in situ} sample and MIM-preparation. Future extensions will include laser post-ionisation to detect also neutral particles as well as a surface barrier detector to obtain complementary data on external electron emission.

\section{Acknowledgement}
We gratefully acknowledge financial support from the DFG within the framework of the SFB 616: Energy dissipation at surfaces. We thank D.~Diesing for many fruitfull discussions and for supplying us with MIMs and his know-how. We thank H.~Lebius and G.~Ban d'Etat for sharing the idea of the ion escalator and W.~Saure for his help with the electronics. Finally we have to thank S. Akc\"oltekin for the preparation of the graphene covered SrTiO$_3$ sample.
\bibliography{iisc17HICS_arXiv}

\begin{thebibliography}{10}

\bibitem{sputterbuch}
R.~Behrisch and W.~Eckstein (eds.).
\newblock Sputtering by particle bombardment.
\newblock {\em Springer Topics in Applied Physics}, 2007.

\bibitem{Zwart1989}
S.~T. de~Zwart, A.~G. Drentje, A.~L. Boers, and R.~Morgenstern.
\newblock {\em Surface Science}, 217(1-2):298--316, 1989.

\bibitem{Elektronenemission}
J.~L{\"o}rin\v{c}\'{\i}k, Z.~\v{S}roubek, H.~Eder, F.~Aumayr, and HP. Winter.
\newblock {\em Phys. Rev. B}, 62:16116, 2000.

\bibitem{SHIreview}
F.~Thibaudau, J.~Cousty, E.~Balanzat, and S.~Bouffard.
\newblock {\em Phys. Rev. Lett.}, 67:1582, 1991.

\bibitem{NatNano}
E.~Akc{\"o}ltekin, T.~Peters, R.~Meyer, A.~Duvenbeck, M.~Klusmann, I.~Monnet,
  H.~Lebius, and M.~Schleberger.
\newblock {\em Nat. Nanotechnol.}, 2:290, 2007.

\bibitem{Fritzreview}
F.~Aumayr, A.S. El-Said, and W.~Meissl.
\newblock {\em Nucl. Instr. Meth. B}, 266:2729, 2008.

\bibitem{HCIreview}
A.~Arnau, F.~Aumayr, P.M. Echenique, M.Grether, W.~Heiland, J.~Limburg,
  R.~Morgenstern, P.~Roncin, S.~Schippers, R.~Schuch, N.~Stolterfoht, P.~Varga,
  T.J.M. Zouros, and HP. Winter.
\newblock {\em Surface Science Reports}, 27:113, 1997.

\bibitem{Werner2001}
T.~Werner, G.~Zschornack, F.~Grossmann, V.P. Ovsyannikov, and F.~Ullmann.
\newblock {\em Nucl. Instr. Meth. B}, 178:260, 2001.

\bibitem{Marrs1988}
R.~E. Marrs, M.~A. Levine, D.~A. Knapp, and J.~R. Henderson.
\newblock {\em Phys. Rev. Lett.}, 60:1715, 1988.

\bibitem{Meyer2004}
S.~Meyer, D.~Diesing, and A.~Wucher.
\newblock {\em Phys. Rev. Lett.}, 93(13):137601, 2004.

\bibitem{Peters2008}
T.~Peters, C.~Haake, D.~Diesing, D.~Kovacs, A.~Golczewski, G.~Kowarik,
  F.~Aumayr, A.~Wucher, and M.~Schleberger.
\newblock {\em New J. Phys}, 10:073019, 2008.

\bibitem{Kovacs2008}
D.A. Kovacs, T.~Peters, C.~Haake, M.~Schleberger, A.~Wucher, A.~Golczewski,
  F.~Aumayr, and D.~Diesing.
\newblock {\em Phys. Rev. B}, 77:245432, 2008.

\bibitem{Wucher1993a}
A.~Wucher, M.~Wahl, and H.~Oechsner.
\newblock {\em Nucl. Instr. Meth. B}, 82:337, 1993.

\bibitem{Graphene}
A.~K. Geim and K.~S. Novoselov.
\newblock {\em Nature Materials}, 6:183, 2007.

\bibitem{SAkcoeltekin08}
S.~Akc{\"o}ltekin, M.~El Kharazzi, B.~K{\"o}hler, A.~Lorke, and M.~Schleberger.
\newblock {\em In preparation}.

\bibitem{ElSaid2008}
A.~S. El-Said, R.~Heller, W.~Meissl, R.~Ritter, S.~Facsko, C.~Lemell,
  B.~Solleder, I.~C. Gebeshuber, G.~Betz, M.~Toulemonde, W.~M{\"o}ller,
  J.~Burgd{\"o}rfer, and F.~Aumayr.
\newblock {\em Phys. Rev. Lett.}, 100(23):237601, 2008.

\bibitem{Khalfaoui2006}
N.~Khalfaoui, M.~G{\"o}rlich, C.~M{\"u}ller, M.~Schleberger, and H.~Lebius.
\newblock {\em Nucl. Instr. Meth. B}, 245:246, 2006.

\bibitem{Akcoeltekin08}
E.~Akc{\"o}ltekin, S.~Akc{\"o}ltekin, O.~Osmani, H.~Lebius, and M.~Schleberger.
\newblock {\em New J. Phys.}, 10:053007, 2008.

\end{thebibliography}

\end{document}